\documentclass[%
 reprint,
superscriptaddress,
 aps,
 pra,
longbibliography,
]{revtex4-2}
\usepackage[latin1]{inputenc}
\usepackage{graphicx}
\usepackage{dcolumn}
\usepackage{bm}
\usepackage{natbib}
\usepackage{amsmath,bm}
\usepackage{xspace}
\usepackage{amssymb}
\usepackage{hyperref}
\usepackage{siunitx}
\usepackage[a4paper,margin=2cm]{geometry}
\usepackage[caption=false]{subfig}
\usepackage{booktabs}
\usepackage[export]{adjustbox}
\usepackage{multirow}
\usepackage{physics}
\usepackage{accents}

\usepackage{multibib}
\usepackage[dvipsnames]{xcolor}

\newcommand{\ahat}{\ensuremath{\hat{a}}\xspace}
\newcommand{\ahatd}{\ensuremath{\hat{a}^{\dagger}}\xspace}
\newcommand{\bhat}{\ensuremath{\hat{b}}\xspace}
\newcommand{\bhatd}{\ensuremath{\hat{b}^{\dagger}}\xspace}

\newcommand{\Supp}[1]{(Appendix #1)} 
\newcommand{\LetterPRL}[0]{Letter}
\newcommand{\SectionPRL}[1]{\emph{#1.---~}}
\begin{document}

\title{Signatures of Quantum Gravity in the Gravitational Self-Interaction of Photons}

\author{Zain Mehdi}
 \email{zain.mehdi@anu.edu.au}
 \affiliation{Department of Quantum Science and Technology and Department of Fundamental and Theoretical Physics, Research School of Physics, Australian National University, Canberra 2600, Australia}%
 \author{Joseph J. Hope}
  \affiliation{Department of Quantum Science and Technology and Department of Fundamental and Theoretical Physics, Research School of Physics, Australian National University, Canberra 2600, Australia}%
 \author{Simon A. Haine}
 \email{simon.haine@anu.edu.au}
 \affiliation{Department of Quantum Science and Technology and Department of Fundamental and Theoretical Physics, Research School of Physics, Australian National University, Canberra 2600, Australia}%

\date{\today}

\begin{abstract}
We propose relativistic tests of quantum gravity using the gravitational self-interaction of photons in a cavity. We demonstrate that this interaction results in a number of quantum gravitational signatures in the quantum state of the light that cannot be reproduced by any classical theory of gravity. We rigorously assess these effects using quantum parameter estimation theory, and discuss simple measurement schemes that optimally extract their signatures. Crucially, the proposed tests are free of QED photon-photon scattering, are sensitive to the spin of the mediating gravitons, and can probe the locality of the gravitational interaction. These protocols provide a new avenue for studying the quantum nature of gravity in a relativistic setting.  
\end{abstract}


\maketitle

\SectionPRL{Introduction}The lack of experimental evidence supporting the quantization of the gravitational field is one of the most significant challenges in the search for a unified theory of quantum theory and general relativity (GR). Remarkably, low-energy experiments may soon be capable of testing the quantum nature of the gravitational field~\cite{Hossenfelder2017,bose_spin_2017,marletto_gravitationally_2017,Krisnanda2017,marletto_when_2018,Pino2018,Howl2017,Carney_2019,carlesso_testing_2019,Christodoulou2019,krisnanda_observable_2020,marshman_locality_2020,van_de_Kamp_2020,haine_searching_2021,howl_non-gaussianity_2021,Weiss2017,Bose2022Mechanism,Anastopoulos2020,Belenchia2016,Derakhshani2016,Schmle2016,AlBalushi2018,Miao2020,Tilly2021,Schut2022,Pedernales2022}. In particular, proposals for observing for quantum correlations between masses induced by their gravitational interaction have seen significant development in the last five years~\cite{marletto_when_2018,Carney_2019,carlesso_testing_2019,Pedernales2022,krisnanda_observable_2020,Christodoulou2019,marshman_locality_2020,van_de_Kamp_2020,Chevalier2020,Rijavec2021,haine_searching_2021,Toros2021,howl_non-gaussianity_2021,Tilly2021,Bose2022Mechanism,Schut2022}. The majority of these schemes consider gravitationally-induced entanglement (GIE) as the key signature of a quantized gravitational field, following proposals by Bose~\emph{et~al.}~\cite{bose_spin_2017} and Marletto and Vedral~\cite{marletto_gravitationally_2017}. More recent proposals have studied further signatures of quantum gravity that may be probed in ultra-cold atomic ensembles~\cite{haine_searching_2021,howl_non-gaussianity_2021} and optomechanical systems~\cite{AlBalushi2018,Miao2020,Matsumura2020, Biswas:2022}.

While the prospect of table-top tests of quantum gravity is extraordinary, there are significant challenges with realising these proposals experimentally. Firstly, there are the well-known technical challenges of creating and probing macroscopic superpositions of mesoscopic masses (${\sim}10^{-14}$kg) in the presence of environmental decoherence~\cite{van_de_Kamp_2020,Rijavec2021}, as well as the requirement that Casimir-Polder interactions between the masses must be suppressed~\cite{Chevalier2020,van_de_Kamp_2020}. Still, these challenges may be overcome within the next decade~\cite{Margalit2021}. There are additional conceptual challenges concerning arguments that experiments that rely on the Newtonian gravitational interaction cannot truly attest to the quantum nature of the gravitational field~\cite{Anastopoulos2018comment,Anastopoulos2020,Anastopoulos2021,Anastopoulos2021b,Martinez2022arxiv,Anastopoulos2022}. These objections to GIE tests of quantum gravity have seen significant debate~\cite{Anastopoulos2018comment,Belenchia2018,Belenchia2019,Christodoulou2019,marshman_locality_2020,Anastopoulos2020,Anastopoulos2021,Anastopoulos2021b,Martinez2022arxiv,Danielson2022,Anastopoulos2022,Carney2022,Huggett2022,Adlam2022} -- motivating the need for tests of quantum gravity beyond the Newtonian regime.\par

In this \LetterPRL, we propose a platform for fundamentally relativistic tests of quantum gravity, using the gravitational self-interaction of quantum light in a cavity. We demonstrate that there are multiple quantum gravitational signatures that can be extracted from the quantum state of the light, that cannot be reproduced by \emph{any} classical theory of gravity. We provide a metrological analysis of the proposed tests, and discuss challenges with their experimental realization. The proposed protocols are sensitive to the spin of the virtual gravitons that mediate the gravitational interaction between photons, are naturally free of QED photon-photon scattering, and can probe retardation effects to test the locality of the gravitational interaction. This work therefore provides a new approach for probing the quantum nature of the gravitational field including relativistic aspects that cannot be probed in the Newtonian regime.

\SectionPRL{Self-interaction of light due to gravity}A sufficiently high-density electromagnetic (EM) field will experience a self-interaction mediated by the gravitational field: the energy-density of light will curve spacetime, which in turn will affect the path of the light. This effect is well studied for freely propagating light, both classically~\cite{Tolman1931,Wheeler1955,Ratzel_2016,vanHolten2018} and for photons~\cite{Barker1967,Westervelt1970,Vilasi2011,Brodin2006,Ratzel2016Entanglement,Lagouvardos2021}. In the case of quantum light, the nature of this self-interaction will depend on whether or not the gravitational field is also quantized -- providing a powerful avenue of probing the quantum nature of gravity with massless photons.

This gravitational self-interaction of light is well described within the framework of perturbative GR, in which the spacetime metric is written as $g_{\mu\nu} \approx \eta_{\mu\nu} + h_{\mu \nu}$ for a background (Minkowski) metric $\eta_{\mu\nu}$ and a perturbation $h_{\mu \nu}$. In the Lorenz gauge, this leads to the linearized Einstein field equations~\cite{Flanagan2005}
\begin{align}
    \label{eq:LinearedFieldEq_EM}
        \Box \bar{h}^{\mu \nu} = -\frac{16\pi G}{c^4} T_{(0)}^{\mu \nu}\,,
    \end{align}
where $\Box=\nabla^2 - \partial_t^2/c^2$, 
$\bar{h}^{\mu \nu} = h^{\mu \nu}-\frac{1}{2}\eta^{\mu \nu} h^{\alpha}_\alpha$, 
and $T_{(0)}^{\mu \nu}$ is the stress-energy tensor of the matter source (the subscript indicates evaluation on the Minkowski background).

Here we consider the gravitational field sourced by an EM field, for which the stress energy tensor is $\mu_0 T_{(0)}^{\mu\nu} =  F^{\mu\lambda}F^\nu_\lambda - \eta^{\mu \nu}F^{\alpha \beta}F_{\alpha\beta}/4$, where $F^{\mu\nu}=\partial^\mu A^\nu - \partial^\nu A^\mu$ is the EM field tensor in terms of the four-potential $A^\mu$, and we have employed Einstein summation. Note that the EM stress energy tensor is traceless $T^{\mu}_{\mu}=0$, which via Eq.~\eqref{eq:LinearedFieldEq_EM} gives $ h^{\alpha}_\alpha=0$. Eq.~\eqref{eq:LinearedFieldEq_EM} then has the retarded Greens function solution
\begin{align}
	h^{\mu \nu}(\bm{r},t) =\frac{4 G}{c^4}\int d\bm{r'}\frac{T^{\mu \nu}_{(0)}\left(\bm{r'},t-\frac{|\bm{r}-\bm{r'}|}{c}\right)}{|\bm{r}-\bm{r'}|} \label{EFsol1} \,.
\end{align}

The gravitational interaction energy can be treated within the linearized gravity framework starting from the interaction Lagrangian density $\mathcal{L}_{\rm int}=h_{\mu\nu}T^{\mu\nu}_{(0)}/2$~\cite{Maggiore2007}, which corresponds to the interaction Hamiltonian density (Appendix A):
\begin{align}
\label{eq:HamDensInt}
	\mathcal{H}_{\rm int}(\bm{r}) = \frac{1}{\mu_0}h^{\mu 0}  F_{\mu\lambda} F_{0}^{\lambda}- \frac{1}{2}h^{\mu\nu}T^{(0)}_{\mu\nu} \,.
\end{align}
The self-interaction Hamiltonian is then obtained by substitution of Eq.~\eqref{EFsol1}
into $\int d^3\bm{r}\mathcal{H}_{\rm int}(\bm{r})$, which is quartic in EM fields, non-local, and contains non-negligible retardation effects in general.

\begin{figure}
	\centering
    \includegraphics[width=.9\columnwidth]{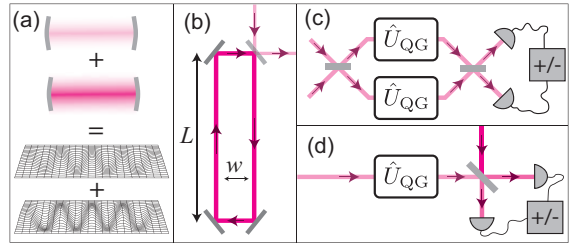}
	\caption{(a) Superpositions of different photon numbers result in superpositions of spacetime geometries if gravity is quantum, resulting in quantum gravitational signatures in the state of the light. (b) Cavity geometry considered, where the dominant gravitational interaction is between the two long arms, $L\gg w$. (c-d) Near-optimal measurement schemes for observing signatures of quantum gravity in (c) the interference between a pair of cavities or (d) non-Gaussianity in the phase quadratures of photons exiting a single cavity. $\hat{U}_{\rm QG}$ represents the system in (b). }  
   \label{fig:numberscheme}
\end{figure}

The quantization of the EM field gives a stress-energy tensor that is operator valued, i.e. $\hat{T}^{\mu \nu}$, which directly implies the quantization of the gravitational field via Eq.~\eqref{eq:LinearedFieldEq_EM}, i.e. $\hat{h}^{\mu \nu}$ (Fig.~\ref{fig:numberscheme}a). 
The self-interaction term in the resulting Hamiltonian will then be quartic in bosonic creation/annihilation operators -- e.g. $\hat{H}_{\rm int}= \int d^3\bm{r}\mathcal{H}_{\rm int}(\bm{r})\sim \hat{a}^\dag\hat{a}^\dag\hat{a}\hat{a}$ for a single-mode field. In contrast, the self-interaction Hamiltonian for any \emph{classical} theory of gravity -- specifically, any theory for which $h^{\mu \nu}$ 
is not operator valued-- may only be quadratic in creation/annihilation operators, at most. For example, in semiclassical gravity where $\hat{T}^{\mu \nu}\to \langle\hat{T}^{\mu \nu}\rangle$ in Eq.~\eqref{EFsol1}, we would have self-interaction Hamiltonian terms such as $\sim\langle\hat{a}^\dag\hat{a}\rangle\hat{a}^\dag\hat{a}$. Therefore there will be distinct signatures of the quantum self-interaction that cannot be reproduced by \emph{any} classical theory of gravity. We will explore these signatures both qualitatively and quantitatively in this work.\par 

Crucially, the self-interaction is sensitive to the spin of the quantized gravitational mediator, unlike Newtonian interactions~\cite{Carney2022}. The self-interaction described by Eq.~\eqref{eq:HamDensInt} specifically requires a spin-$2$ graviton -- mediators with any other spin would correspond to a different tensorial structure to the gravity-light coupling, and thus give rise to a distinct interaction from $\hat{H}_{\rm int}$~\cite{Carney2022}. As the EM field has a traceless stress-energy tensor ($T=T^{\mu}_\mu=0$), this rules out mediation by spin-$0$ gravitons that couple solely to the trace $T$. This includes virtual spin-$0$ gravitons that appear in the canonical quantization of perturbative GR~\cite{Gupta_1952}, and quantized scalar gravity theories, such as that of Nordstr\"om~\cite{Nordstrm1913,Carney2022} or Brans-Dicke theory~\cite{Brans1961,Banerjee2016,Pal2016}.
We note here the related proposals for realising the quantum analog of gravitational light-bending~\cite{Biswas:2022}, and the photonic analog of the GIE tests~\cite{Aimet2022_Preprint}; which can similarly discern between mediators of different spin.    \par 

\SectionPRL{Evading QED photon-photon scattering} At large EM field densities, photons can also interact by exchange of virtual electron-positron pairs as described by quantum electrodynamics (QED); for the optical frequencies we consider here this is typically many orders of magnitude stronger than the gravitational interaction~\cite{Brodin2006}. To avoid this problem, we exploit two key properties of the QED photon-photon interaction: it vanishes for monochromatic plane waves~\cite{Schwinger1951,Marklund2004}, and is a local interaction. In contrast, the quantum gravitational self-interaction is \emph{non-local}, and thus permits non-zero interactions in geometries where there is no QED self-interaction. \par 
Here we consider a travelling-wave (ring) cavity of finesse $\mathcal{F}$ in a rectangular geometry (Fig.~\ref{fig:numberscheme}b), such that there is a gravitational self-interaction between spatially-separated parts of the cavity mode. Specifically, we consider the case where the dominant interaction is between the counter-propagating light in the long arms each of length $L$, separated by a small distance $w\ll L$ -- the quantum analog of Ref.~\cite{Kramer1998}. 
An appealing aspect of this system is that, by choice of optical polarization, both static and non-static gravitational perturbations can be studied. The latter case, where retardation effects will be significant, can explicitly test the locality of the interaction; a key aspect of quantum gravity inaccessible to tests with non-relativistic sources~\cite{Christodoulou2022}. In this work we will focus on the case of circularly polarized light, for which $T^{\mu\nu}$ is static, however the qualitative features of the interaction will be unchanged for general polarizations.

\SectionPRL{Quantum gravitational Kerr effect}First, let us consider the single-mode self-interaction of a cavity field of frequency $\omega_0$ described by creation/annihilation operators $\hat{a}$/$\hat{a}^\dag$. We will consider the multi-mode case where vacuum modes are included, shortly. 
Assuming the two long arms of the cavity are well separated such that diffraction effects can be neglected ($w\gg \lambda$), the leading contribution to the self-interaction Hamiltonian in the limit $L\gg w$ is given by (Appendix B):
\begin{align}
\label{eq:HQG_SM}	\hat{H}_{\rm QG} = -\frac{16G}{L}\left(\frac{\hbar\omega_0}{c^2}\right)^2\log\left(\frac{L}{w}\right)\hat{a}^\dag\hat{a}^\dag\hat{a}\hat{a}\,.
\end{align} 
Applying this interaction Hamiltonian for the cavity interrogation time $\tau\approx 2\mathcal{F}L/c$ (assuming $L\gg w$) generates the unitary $\hat{U}_{\rm QG} = \exp\left(i\chi_{\rm Q} \hat{a}^\dag\hat{a}^\dag\hat{a}\hat{a}\right)$, where 
\begin{align} 
\label{eq:chiQ}
\chi_{\rm Q} \approx \frac{32G\mathcal{F}\hbar\omega_0^2}{c^5}\log\left(\frac{L}{w}\right)\,.
\end{align}
Eq.~\eqref{eq:HQG_SM} can be understood as a gravitational Kerr effect where the graviton vacuum plays the role of the underlying nonlinear medium.  
Interactions of the form Eq.~\eqref{eq:HQG_SM} will produce entanglement between photons, which means that observing signatures of the quantum gravitation self-interaction will also be indirect witnesses of GIE between photons.

In contrast, consider a general classical theory of gravity for which $h^{\mu \nu}$ is not operator-valued, which will result in a self-interaction term of quadratic order. 
In a rotating-wave approximation (RWA)~\footnote{The validity of an RWA is justified here by the long cavity interrogation time, $\tau\gg\omega_0^{-1}$.}, the classical gravitational Hamiltonian must then take the form $\hat{H}^{\rm C}_{\rm int}=\hbar\lambda_{\rm C} \hat{a}^\dag \hat{a}$ for some coefficient $\lambda_{\rm C}$, giving the unitary $\hat{U}_{\rm C} = \exp\left(i\chi_{\rm C} \hat{a}^\dag\hat{a}\right)$, where $\chi_{\rm C}=\lambda_{\rm C}\tau$. Note that $\lambda_{\rm C}$ itself may depend on expectations of the quantum field as in semi-classical gravity, and may even include contributions from non-gravitational effects that do not result in interactions between photons. Nevertheless, any such Hamiltonian cannot reproduce features of the quantum gravitational Kerr effect. \par

\SectionPRL{Fundamental detectability}Here we consider the fundamental limit to which the quantum gravitational self-interaction can be distinguished from the self-interaction described by any classical theory of gravity. More precisely, we ask the question -- how well can $\chi_{\rm Q}$ be distinguished from zero, given a completely \emph{unknown} value of $\chi_{\rm C}$? We can reformulate this as a multi-parameter estimation problem by treating $\chi_{\rm C}$ as a \emph{nuisance parameter}, following the approach of Ref.~\cite{haine_searching_2021}. Specifically, the smallest possible non-zero value of $\chi_{\rm Q}$ that may be inferred from a set of $M$ measurements each with interrogation time $\tau$ 
is given by the quantum Cram\'er-Rao bound (QCRB) \cite{Liu2019}: 
\begin{align}\label{eq:QCRB}\chi_{\rm Q} \geq \sqrt{[F_{i,j}]^{-1}_{\rm Q,Q}/M}\,,\end{align} 
where
$	F_{i,j} = 4\tau^2\,\textrm{Cov}(\hat{G}_i,\hat{G}_j)$
is the quantum Fisher information matrix (QFIM) with respect to the classical and quantum generators $\hat{G}_{\rm C}=\hat{a}^\dag\hat{a}$ and $\hat{G}_{\rm Q}=\hat{a}^\dag\hat{a}^\dag\hat{a}\hat{a}$, respectively. The covariance is computed with respect to the state $\ket{\Psi}=\hat{U}_{\rm QG}\hat{U}_{\rm C}\ket{\psi}$ for some initial state $\ket{\psi}$. Note that the existence of an inverse QFIM is not guaranteed -- a non-invertible QFIM would correspond to the effect of the quantum gravitational interaction being fundamentally indistinguishable from a classical gravitational interaction.  

The choice of initial state $\ket{\psi}$ is therefore crucial to the detectability of quantum gravity given ignorance of $\chi_{\rm C} $; the wrong choice of state may result in zero sensitivity to $\chi_{\rm Q}$ regardless of the number of measurements. In this analysis we restrict ourselves to quantum states that may be prepared with established experimental techniques. In particular, we find that for a squeezed vacuum state, the signatures of quantum and classical gravity theories are highly distinguishable, resulting in Heisenberg scaling of the sensitivity~\Supp{D} 
\begin{align}
\label{eq:QCRB_SqVac} \chi_{\rm Q}\geq [ {96M} N(N+1)]^{-1}
\end{align}
where $N\equiv \langle \hat{a}^\dag \hat{a} \rangle$ is the average number of photons in the cavity. In comparison, a significant amount of information is lost for coherent state or a squeezed coherent state, with the sensitivity scaling as $N^{-3/2}$ in the absence of the nuisance parameter, and as $N^{-1}$ in its presence. \par 

This result allows us to make strong statements about the fundamental detectability of $\chi_{\rm Q}$, given no information about the classical theories of gravity from which this signal is distinguished. Specifically, we can use Eq.~\eqref{eq:QCRB_SqVac} to bound the experimental requirements to infer a non-zero value of $\chi_{\rm Q}$. For example, we may re-write Eq.~\eqref{eq:chiQ} in terms of the circulating power in the cavity $P_{\rm circ}\approx N\hbar\omega_0 c/(2L)$, and the number of experiments  $M\approx T/\tau$ that may be conducted within a total time $T$. For $N\gg 1$, 
we find that the circulating power required to satisfy Eq.~\eqref{eq:QCRB_SqVac} is approximately:
\begin{align}
	P_{\rm circ} \gtrsim \frac{c^3}{16}\left(\frac{c\hbar^2}{12 G^2\mathcal{F} L^3 T \log(L/w)^2}\right)^{1/4} \,.
\end{align} 
This expression shows weak dependence 
on the interrogation period $P_{\rm circ}\sim T^{-1/4}$, and is most strongly dependent on the cavity length $L$. 
As an example, consider experimental parameters similar to those expected of the next-generation gravitational-wave detector Cosmic Explorer~\cite{Abbott_2017}: a $L=10$km long cavity of finesse $\mathcal{F}=450$ with a $2\mu$m wavelength laser and long arm separation $w=10$cm. The minimum required circulating power for this system to infer the quantum signature of gravity over a year-long interrogation period is roughly $125$MW -- much larger than the $\mathcal{O}(1)$MW circulating powers expected of Cosmic Explorer. With a moderate increase to the finesse ($\mathcal{F}=10^3$), the required circulating power could in-principle be achieved with a pump laser power of roughly $100$kW. Interestingly, this level of laser power is significantly lower than quoted requirements for directly measuring the gravitational field of the light~\cite{Spengler2022}, or generating observable light-matter entanglement~\cite{Biswas:2022}. The weak logarithmic dependence of the interaction on the separation $w$ means the beam width $\sigma$ can be made large enough to prevent damage to the mirror surfaces while satisfying $\sigma\ll w$: e.g. a beam width of $2$cm corresponds to an intensity of roughly $25$MW/cm$^2$ for the above laser power, which is well within the thermal tolerance of modern low-loss cavity mirrors~\cite{Meng2005}. While $100$kW continuous lasers are currently available~\cite{ipg,Shcherbakov2013}, the requirement of a squeezed-vacuum state at this power is beyond current experiments -- we will see later the need for highly squeezed states may be avoided by considering multi-mode signatures of the gravitational interaction.

\SectionPRL{Possible measurement schemes}
The question remains to design a measurement scheme that optimally extracts the signature of quantum gravity in the proposed cavity system, i.e. a measurement scheme for which the ideal precision is given by the right hand side of Eq.~\eqref{eq:QCRB_SqVac}. Optimal measurement schemes are not always guaranteed to exist, and are often not experimentally feasible. We instead identify two, experimentally realizable, \emph{nearly} optimal schemes -- their precision is not exactly Eq.~\eqref{eq:QCRB_SqVac}, but scales the same with $N$ (Heisenberg scaling).

In Figure~\ref{fig:numberscheme}(c) we present a simple interferometric scheme that provides an optimal measurement for an initial two-mode squeezed vacuum state (TMSV)~\Supp{E}. In this scheme, the initial TMSV is beam-split to create a pair of squeezed vacuum states, each of which is fed into a high-finesse cavity. The gravitational self-interaction of the light leads to a relative phase between the two modes, which is then read-out by interfering the light exiting the cavity and reading out the number difference between the two modes. The distinguishability of the signature of quantum gravity from classical gravity in this scheme can be simply understood by considering the evolution of the state in the number basis. For the initial TMSV state $\ket{\psi} = \sum_N c_N \ket{N,N}$, where $c_N=(-e^{i\phi}\tanh(r))^N/\cosh(r)$, the measurement scheme can be described by the unitary operator $\hat{U}_{\rm MZ}=\hat{U}_{\rm BS}(\pi/2)\hat{U}_{\rm QG }\hat{U}_{\rm BS}(\pi/2)$, where $\hat{U}_{\rm BS}(\pi/2)$ is a 50-50 beamsplitter and $\hat{U}_{\rm QG }=\exp\left(i\chi_{\rm Q}((\hat{a}^\dag)^2\hat{a}^2+(\hat{b}^\dag)^2\hat{b}^2)\right)$ is the quantum-gravitational self-interaction for the two cavity modes represented by $\hat{a}$ and $\hat{b}$. To first order in $\chi_{\rm Q}\ll 1$, $\hat{U}_{\rm MZ}$ couples each $\ket{N,N}$ in the initial TMSV to $\ket{N-2,N+2}$ and $\ket{N+2,N-2}$. Higher order terms in this expansion will result in coupling to states $\ket{N-2n,N+2n}$ for integer $n$, though these will be negligible due to the exceptionally small value of $\chi_{\rm Q}$. Number measurements on each mode will then collapse the quantum state onto a manifold of well-defined $N$, with a small probability $\sim\chi_{\rm Q}^2$ of measuring a non-zero number difference between the modes. 

In contrast, consider the same scheme for a classical theory of gravity. First we assume the two cavities experience precisely the same self-interaction -- i.e. $H_{\rm int}^{\rm C}=\hbar\lambda_{\rm C}\left(\hat{a}^\dag\hat{a}+\hat{b}^\dag\hat{b}\right)$ -- the number distribution of each mode measured at the detector is unchanged from the number distribution of the original TMSV state. More generally, we allow for the case of a classical theory of gravity that additionally has a term of the form $H_{\rm int}^{\rm C}\sim\hat{a}^\dag\hat{a}-\hat{b}^\dag\hat{b}$. Such a term could also be contributed by non-gravitational effects that result in an asymmetry between the two cavities. In this case, each $\ket{N,N}$ in the initial TMSV will be coupled to states $\ket{N-n,N+n}$ for integer $n$ by the measurement scheme, with dominant coupling for $n=\pm 1$. This is clearly distinguishable from the signature of quantum gravity, in which the dominant coupling is for $n=\pm 2$.  The main limitation with this approach is the requirement of ultra-high efficiency detectors operating at the single photon level, which for the high-power required is beyond reason for modern detection technology.

An appealing alternative measurement scheme uses homodyne detection to observe signatures of quantum gravity in high-order cumulants of the phase quadratures \cite{howl_non-gaussianity_2021}, which can be implemented straightforwardly with a single cavity and homodyne detection (Fig.~\ref{fig:numberscheme}c). This approach relies on the fact that the quartic self-interaction term described by a quantum theory of gravity will generate non-Gaussianity of the optical field, that cannot be reproduced by \emph{any} classical theory of gravity \cite{howl_non-gaussianity_2021}. As a result, a non-zero value of the fourth-order cumulant in a phase-quadrature of the light is a clear signature of the quantum nature of gravity. This scheme was initially proposed in the context of massive ultra-cold atomic systems \cite{howl_non-gaussianity_2021}, in which observing this signature is considerably more challenging due to the inability to directly measure the phase quadratures of massive particles. For this scheme it will again be optimal to use a high-power squeezed vacuum state of light, to maximize the sensitivity of the state to the gravitational self-interaction \cite{howl_non-gaussianity_2021}.

While our proposed tests are free of QED photon-photon interactions, radiation pressure on the cavity mirrors can also result in a quantum interaction between photons. Fortunately, radiation pressure can be reduced by rigidly coupling the mirrors to the Earth -- seismic noise generates only a classical phase shift and does not mimic signatures of photon-photon interactions. However, there may be other challenges associated with microscopic material surface effects at high laser intensities that are poorly understood, and merit further investigation in the context of the proposed tests.

\SectionPRL{Third harmonic generation} In addition to the gravitational Kerr non-linearity, there are additional signatures of quantum gravity that arise when we include multiple modes of the quantized optical field in our calculation. Specifically, we may have a four-wave mixing process between photons in the pump mode ($\omega_0$) that can populate higher frequency  resonances $n\omega_0$, where $n\in [2,3,\dots]$. For our system, where only a single frequency is initially populated, the dominant four-wave mixing process is third-harmonic generation, i.e. $H_{\rm int}\sim \hat{a}_{\omega_0}\hat{a}_{\omega_0}\hat{a}_{\omega_0}\hat{a}_{3\omega_0}^\dag$. In contrast, any classical gravitational theories can only have non-energy-conserving terms arising such as $ H_{\rm int}\sim \hat{a}_{\omega_0}\hat{a}_{3\omega_0}^\dag$ which in the interaction picture will oscillate with frequency $2\omega_0$, and can thus be discarded in a RWA. Therefore, the detection of photons of frequency $3\omega_0$ exiting the cavity would be a clear signature of quantum gravity. This could in-principle be achieved with modern photodetectors that operate at the single-photon level, by discarding the pump mode using a wave-length selective beamsplitter. The key benefit of this experiment is that it can be implemented with coherent light -- with the QCRB for the sensitivity scaling as $N^{-3/2}$ for both coherent and squeezed vacuum states of the pump mode. 

\SectionPRL{Conclusions} Near-future experiments seeking to observe GIE between masses may soon provide the first evidence to support the quantum nature of gravity. However, such non-relativistic experiments can only probe the quantization of the Newtonian interaction and cannot attest to other aspects of quantum gravity using quantum light, such as the locality of the gravitational interaction or the spin of the graviton. This work may provide a new pathway for probing relativistic aspects of quantum gravity; together with near-term GIE experiments the proposed tests may provide a more detailed insight to the true nature of the gravitational field. \par 

While the proposed tests provide a powerful avenue to probing quantum gravity, they will certainly require improvements to current experimental capabilities. In particular, reading out signatures of quantum gravity from photon statistics will require photon detection efficiencies well beyond current experimental capabilities -- though there are clear routes to reducing these requirements by using interaction-based readouts~\cite{Nolan:2017b,Haine2018}, for example. Reducing technological requirements should be the focus of future investigations that may, for example, consider alternate measurement schemes or more complicated geometries. Regardless, we expect the experimental requirements will become increasingly realistic in the future, given the outstanding rate at which photonic quantum technologies are advancing. In the near term, the experiments proposed here can also be adapted to be precision tests of QED, by using a standing-wave of light rather than a travelling wave, such that the QED photon-photon interaction is non-vanishing. \par

\SectionPRL{Acknowledgements}We acknowledge insightful discussions with Daniel Carney, Ruvi Lecamwasam, James Gardner, and Karl Wette. ZM is supported by an Australian Government Research Training Program (RTP) Scholarship. SH acknowledges support through an Australian Research Council Future Fellowship grant FT210100809. 
\pagebreak
\widetext
\appendix 
\section{Derivation of the gravitational interaction Hamiltonian \label{app:DerivationFP}}
Here we provide a detailed derivation of the self-interaction Hamiltonian $\hat{H}_{\rm int}$ given in the main text. We start with the complete action describing the coupled light-gravity system:
\begin{align}
	S &= S_{\rm EM} + S_{\rm grav} \,.
\end{align}
Here $ S_{\rm EM}$ is the Maxwell action
\begin{align}
	S_{\rm EM}&=-\frac{1}{4\mu_0}\int d^4x\sqrt{-g}F_{\mu\nu}F^{\mu\nu} \,,
\end{align}
where $F_{\mu\nu} = \partial_\mu A_\nu - \partial_{\nu} A_{\mu}$ is the EM field tensor, and $g\equiv {\rm det}(g_{\mu\nu})$. The Einstein-Hilbert action $S_{\rm grav}$ is given by:
\begin{align}
	S_{\rm grav} &= \frac{1}{16\pi G}\int d^4x\sqrt{-g}R \,,
\end{align}
where $R\equiv g^{\mu\nu}R_{\mu\nu}$ is the Ricci scalar. We proceed by expanding the gravitational metric around a Minkowski background $g_{\mu\nu}=\eta_{\mu\nu}+h_{\mu\nu}$, in the $[-,+,+,+]$ convention. Next, we expand the Lagrangian density of the Maxwell action to linear order in the gravitational perturbation $h_{\mu\nu}$:
\begin{align}
	\mathcal{L}_{\rm EM}&=-\frac{1}{4\mu_0}F_{\mu\nu}g^{\alpha\mu}g^{\beta\nu}F^{\alpha\beta} +\mathcal{O}(h^2) \\
	&= \mathcal{L}_{(0)} + \mathcal{L}_{(1)} + \mathcal{O}(h^2)\,,
\end{align}
where we have decomposed the Lagrangian density into a free (flat-spacetime) component $\mathcal{L}_{(0)} = -F_{\mu\nu}F^{\mu\nu}/(4\mu_0)$, and the first order correction
\begin{align}
	\mathcal{L}_{(1)} &= -\frac{h}{8\mu_0}F_{\mu\nu}F^{\mu\nu}  - \frac{1}{2\mu_0}h^{\mu\beta}\eta^{\nu \lambda}F_{\lambda\beta} F_{\mu\nu} \,,\\
	&= \frac{1}{2}h^{\mu\nu}T^{(0)}_{\mu\nu}\label{eq:intLag} \,,
\end{align}
where $h \equiv \eta^{\mu\nu}h_{\mu\nu}$, $T^{(0)}_{\mu\nu}$ is the stress-energy tensor of the free EM field on a Minkowski background:
\begin{align}
\label{eq:FlatSpaceStressEnergy}
	T^{(0)}_{\mu\nu} = \frac{1}{\mu_0}\left( F_{\mu\lambda}F_\nu^\lambda - \frac{1}{4}\eta_{\mu \nu}F^{\alpha \beta}F_{\alpha\beta}\right) \,,
\end{align}
noting the symmetry properties $h^{\alpha\beta}=h^{\beta\alpha}$ and $F_{\mu\nu}=-F_{\nu\mu}$. This interaction can alternatively be derived by a variational argument, and is the standard linear coupling term between matter and the linearized gravitational field~\cite{Maggiore2007,Carroll2004}.

We then obtain the corresponding interaction Hamiltonian via a Legendre transform of the linearized Lagrangian. As the interaction depends on derivatives of the EM field via $T_{\mu\nu}$, the resulting interaction term is not simply $-\mathcal{L}_{(1)}$ and contains additional terms due to corrections to the conjugate momenta of the EM field:
\begin{align}
	\Pi^\alpha \equiv \frac{\partial \mathcal{L}}{\partial(\partial_0 A_\alpha)} = \Pi^\alpha_{(0)} + \Pi^\alpha_{(1)} \,,
\end{align}
where $\Pi^\alpha_{(0)}$ is the free field result, and 
\begin{align}
	\Pi^\alpha_{(1)} &= \frac{1}{2} h^{\mu\nu}\frac{\partial T_{\mu\nu}^{(0)}}{\partial(\partial_0 A_\alpha)}\\
	& = \frac{1}{\mu_0}\left(\frac{h}{2}F^{\alpha 0}+ h^{\mu 0}\eta^{\alpha\lambda}   F_{\mu\lambda}  - h^{\mu\alpha}\eta^{0\lambda}   F_{\mu\lambda}           \right) \notag
\end{align}
is the first-order correction due to the interaction with the gravitational field.  Then, working in the interaction picture, the interaction Hamiltonian density is given by:
\begin{align}
	\mathcal{H}_{(1)}(\bm{r}) = \Pi^\alpha_{(1)} \partial_0 A_{\alpha} - \mathcal{L}_{(1)} \,.
\end{align}
For the system under consideration, we have $h=0$ by virtue of the vanishing trace of the EM stress energy tensor. Furthermore, for the gravitational field generated by a travelling wave we have 
\begin{align}
	h^{\mu\alpha} F_{\mu 0} = h^{i \alpha} E_{i} =0 \,,
\end{align}
  as a consequence of the explicit form of the gravitational field solution -- $h^{i \alpha}$ only has non-zero components longitudinal to the direction of propagation (see Eq.~\eqref{eq:hmatrix_LR}), to which the electric field $E_{i}$ is transverse. Therefore for our system the interaction Hamiltonian density can be simplified to (noting $\partial_{0} A_{\alpha} = F_{0\alpha}$):
\begin{align}
	\mathcal{H}_{(1)}(\bm{r}) &= \frac{1}{\mu_0}h^{\mu 0}\eta^{\alpha\lambda}   F_{\mu\lambda} F_{0\alpha}- \frac{1}{2}h^{\mu\nu}T^{(0)}_{\mu\nu} \,.
\end{align}

\section{Lack of gravitational interaction for a free plane wave}
For a monochromatic EM field freely propagating along the $z$ axis, the stress-energy tensor takes the form:
\begin{align}
\label{eq:SETensor_FreePropz}
    T^{\mu \nu}(\bm{r},t) = \mathcal{H}_{(0)}(\bm{r},t)\begin{pmatrix}
    1&0&0&1 \\
    0&0&0&0\\
    0&0&0&0\\
    1&0&0&1 \\
    \end{pmatrix} \,,
\end{align} 
where $\mathcal{H}_{(0)}=(E^2+B^2)/2$ is the (non-interacting) energy-density of the EM field. We will also use the explicit form of the electromagnetic field tensor (keeping the polarization general, for now)
\begin{align}
\label{eq:EMTensor_FreePropz}
	F_{\mu\nu}=\begin{pmatrix}
 0 & -E_x & -E_y & 0 \\
 E_x & 0 & 0 & -B_y \\
 E_y & 0 & 0 & B_x \\
 0 & B_y & -B_x & 0 
\end{pmatrix} \,,
\end{align}
where we have moved to natural units $\mu_0=c=1$.

Firstly, we show that there is no self-interaction for the travelling wave if it is freely propagating. We show this for the case of linearly polarized light, though the calculation is identical for arbitrary polarization. First, from Eq.~\eqref{eq:SETensor_FreePropz} the only non-zero components of the gravitational field are $h_{00}=h_{33}=-h_{30}=-h_{03}$ via the linearized field equation (Eq.~(1) of the main text) -- the minus sign on the off-diagonal terms arises due to lowering indices with the Minkowski metric. Substituting this result, Eq.~\eqref{eq:EMTensor_FreePropz} and $E_x=cB_y$, it is straightforward to see that each term in the self-interaction Hamiltonian density (Eq.~(3) of the main text) is precisely zero. This is consistent with the more detailed calculations that show that co-propagating parallel beams of light do not gravitationally interact~\cite{Tolman1931}; we can apply this intuition here by considering a single beam of light as the limit of two parallel beams brought close together, ignoring diffraction effects.

\section{Self-interaction Hamiltonian for rectangular ring-cavity setup \label{app:IntHamDeriv_Ring}}
For the rectangular ring-cavity setup described in the main text, the dominant contribution to the gravitational self-interaction energy will be the interaction between the two long arms, provided the longer arms of the cavity are significantly longer than their separation $L\gg w$ (where $w$ is the length of the shorter arms). Labelling the two long arms as $L$ and $R$, we can then approximately decompose the stress-energy tensor as $T^{\mu\nu}\approx T^{\mu\nu}_L+T^{\mu\nu}_R$, ignoring the contributions from the short arms. The non-zero elements of these tensors are $T^{00}_{L} = T^{33}_{L}= T^{03}_{L} = T^{30}_{L}=\mathcal{H}_{(0)}$ and $T^{00}_{R} = T^{33}_{R}= -T^{03}_{R} = -T^{30}_{R}=\mathcal{H}_{(0)}$, taking the light along the left arm to be travelling in the positive $z$ direction. By linearity of Eq.~\eqref{eq:LinearedFieldEq_EM}, its solution can then be written as
\begin{align}
 h^{\mu\nu} &= h_L^{\mu \nu} + h_R^{\mu \nu}
\end{align} 	
where 
\begin{align}
\label{eq:hmatrix_LR}h_{L,R}^{\mu \nu} &= h_{p}(\bm{r})\begin{pmatrix}
    1&0&0&\pm 1 \\
    0&0&0&0\\
    0&0&0&0\\
    \pm 1&0&0&1 \\
    \end{pmatrix} ,  \\
	h_{p}(\bm{r}) &=\frac{4G}{c^4} \int_{L,R} d^3\bm{r'} \frac{\mathcal{H}_{(0)}(\bm{r'})}{|\bm{r-r'}|} \,. \label{eq:hLR}
\end{align}
Here the notation $\int_{L,R}$ denotes integrating over the left or right arm of the cavity, respectively. Note we have considered here a time-independent Hamiltonian density corresponding to a circular polarization of light for simplicity, though this can be simply extended to the time-dependent case (linear or elliptical polarizations).

Then, noting each arm does not self-interact $T^{\mu\nu}_i h_{\mu\nu}^i=0$ for $i=L,R$~\Supp{C}, we have:
\begin{align}
	- \frac{1}{2}\int d^3\bm{r} h^{\mu\nu}(\bm{r})T^{(0)}_{\mu\nu}(\bm{r} ) &\approx - \frac{1}{2}\int d^3\bm{r}\left( h
	_L^{\mu\nu}(\bm{r})T^{R}_{\mu\nu}(\bm{r} )+h_R^{\mu\nu}(\bm{r})T^{L}_{\mu\nu}(\bm{r} )\right) \\
	&= -2\sum_{i=L,R}\int_i d^3\bm{r} h_i(\bm{r} ) \mathcal{H}_{(0)}(\bm{r} ) \,. \label{eq:hmuTmu_expansion_circpol}
\end{align}

A similar calculation follows for the other term in the interaction Hamiltonian density, $h^{\mu 0}  F_{\mu\lambda} F_{0}^{\lambda}$, where we have to explicitly consider the form of the electric and magnetic fields:
\begin{align}
	\bm{E} \approx \bm{E}_L[\mathbf{e_{z}}] + \bm{E}_R[\mathbf{-e_{z}}] \,, \bm{B} \approx \bm{B}_L[\mathbf{e_{z}}] + \bm{B}_R[\mathbf{-e_{z}}] \,, \notag
\end{align}
where the argument $[\mathbf{n}]$ indicates direction of propagation in terms of a normalized vector $\mathbf{n}$, with $\mathbf{e_i}$ being the unit vectors for $i=x,y,z$. For circularly polarized light, the EM fields are of the form:
\begin{align}
	\bm{E}_L&=\{E_L^x,E_L^y,0 \}\,, \bm{B}_L = \{-E_L^y/c,E_L^x/c,0\} \,, \\
	\bm{E}_R&=\{E_R^x,E_R^y,0 \}\,, \bm{B}_R = \{E_R^y/c,-E_R^x/c,0\} \,,
\end{align}
which, using $\mathcal{H}_{(0)}=(|E|^2+|B|^2)/2$, gives
\begin{align}
\int d^3\bm{r} h^{\mu 0}  F_{\mu\lambda} F_{0}^{\lambda} \approx -2\sum_{i=L,R}\int_i d^3\bm{r} h_i(\bm{r} ) \mathcal{H}_{(0)}(\bm{r} ) \label{eq:Hintterm2_expansion_circpol}.
\end{align}
This is precisely the same expression as Eq.~\eqref{eq:hmuTmu_expansion_circpol}, though in general (e.g. for other polarizations) the two may differ.

Combining Eqs.~\eqref{eq:hmuTmu_expansion_circpol}, \eqref{eq:Hintterm2_expansion_circpol}, and \eqref{eq:hLR} we find the self-interaction energy to be given by the Hamiltonian:
\begin{align} 
	H_{\rm int} \approx 
	 &-16 G\bigg(\int_L d^3\bm{r}\int_R d^3\bm{r'}  \frac{\mathcal{H}_{(0)}(\bm{r})\mathcal{H}_{(0)}(\bm{r'} ) }{|\bm{r-r'}|} +\int_R d^3\bm{r}\int_L d^3\bm{r'}  \frac{\mathcal{H}_{(0)}(\bm{r})\mathcal{H}_{(0)}(\bm{r'} ) }{|\bm{r-r'}|} \bigg) \,. \label{eq:IntHamiltonian_NewtonianForm}
\end{align}
Eq.~\eqref{eq:HQG_SM} can then be derived by expanding the energy-density in terms of photon creation-annihilation operators. As the energy density for our system is constant, it can be expressed in the simple form 
\begin{align}
\mathcal{H}_{(0)}(\bm{r}) &= \left(\frac{\hbar\omega_0}{c^2}\right) |f(\bm{r})|^2 \hat{a}^\dag\hat{a}
\end{align}
where $f(\bm{r})$ is the spatial mode function and we have neglected the zero-point energy (as we are only concerned with the quartic operator term). Provided the two long sides of the cavity are separated by a distance $w$ satisfying $\lambda \ll w \ll L$, where $\lambda=c/(2\pi\omega_0)$ is the optical wavelength, the transverse shape of the field is well approximated by a delta function:
\begin{align}
|f(\bm{r})|^2 &\approx \frac{1}{2(L+w)} \Pi_0^L(z)\delta(y)\left(\delta(x) + \delta(x-w)\right) \,.
\end{align}
Here we have chosen to align the short axis of the cavity with the $x$ axis, such that the bottom left mirror is placed at the origin, and defined the top hat function: $\Pi_0^L(z)=1$ if $0\leq z \leq L$ and zero otherwise. Note the normalization $2(L+w)$ ensures that the mode is appropriately normalized when integrating over the whole cavity. Defining $\rho_0$ as $\rho_0 |f(\bm{r})|^2 = \mathcal{H}_{(0)}(\bm{r})$, and substituting the above expressions into Eq.~\eqref{eq:IntHamiltonian_NewtonianForm}, and exploiting the symmetry of the system, the self-interaction energy is then given by:
\begin{align} 
	H_{\rm int} &\approx  -32G \rho_0^2 \int_{0}^l \int_0^l \frac{1}{\sqrt{(z-z')^2 + w^2}}dz dz' \\
	&=-32G \rho_0^2\bigg(2\left(w-\sqrt{l^2+w^2}\right) 
 + l\log\left(\frac{\sqrt{l^2+w^2}+l}{\sqrt{l^2+w^2}-l}\right)\bigg) \,,\\
	&\approx -64G\rho_0^2 L \log\left(\frac{L}{w}\right) \,,
\end{align}
where in the last line we have taken the limit $L\gg w$. Then, substituting in 
\begin{align}
	\rho_0 &= \left(\frac{\hbar\omega_0}{c^2}\right)\left(\frac{1}{2(L+w)}\right) \hat{a}^\dag\hat{a} \,,
\end{align}
we arrive at the expression for the single-mode self-interaction Hamiltonian,
\begin{align}
\hat{H}_{\rm int} &\approx -16G\left(\frac{\hbar\omega_0}{c^2}\right)^2\left(\frac{1}{L}\right)\log\left(\frac{L}{w}\right)\hat{a}^\dag\hat{a}^\dag\hat{a}\hat{a} \,,
\end{align}
where we have normally ordered the creation and annihilation operators and discarded the quadratic $\hat{a}^\dag \hat{a}$ term which does not contribute to the quantum gravitational signatures outlined in the main text.

\section{Exemplary QFIM calculation}
In the main text, we compute bounds on the achievable sensitivity with which the quantum gravitational signatures can be extracted, based on calculations of the quantum Fisher information matrix (QFIM), for several quantum states. These calculations are straightforward, however it is nonetheless informative to provide the full details for the single-mode gravitational Kerr effect on a squeezed vacuum state.

A squeezed vacuum state may be written as:
\begin{align}
    \ket{\rm SQV} &= \hat{U}_{\rm sq}(\chi)\ket{0} \,,
\end{align}
where $\hat{U}_{\rm sq}(\chi) = e^{(\chi^*\hat{a}\hat{a}-\chi\hat{a}^\dag\hat{a}^\dag)/2}$ is the squeezing unitary operator, and $\chi=re^{i\theta}$ gives the angle $\theta$ and magnitude $r$ of squeezing. Our calculation is greatly simplified by the relation:
\begin{align}
    \hat{U}_{\rm sq}(\chi) f(\hat{a},\hat{a}^\dag) \hat{U}_{\rm sq}(\chi)^\dag &= f(\hat{a}\cosh(r)- \hat{a}^\dag\sinh(r)e^{i\theta},\hat{a}^\dag\cosh(r) -\hat{a}\sinh(r)e^{-i\theta}) \,,
\end{align}
where $f$ is some polynomial function. Calculations involving coherent states can be performed similarly using: $\bra{\alpha}:f(\hat{a},\hat{a}^\dag):\ket{\alpha}= :f(\alpha,\alpha^*):$, where $:\star :$ denotes normal ordering. We may compute arbitrary expectation values applying the above relation and normally ordering the resulting operators. This allows us to compute the QFIM:
\begin{align}
    \mathbf{F}_{\rm Q} &= 4\begin{pmatrix}
   {\rm Var}(\hat{G}_{\rm Q})       & {\rm Covar}(\hat{G}_{\rm Q},\hat{G}_{\rm C})  \\
    {\rm Covar}(\hat{G}_{\rm C},\hat{G}_{\rm Q})        & {\rm Var}(\hat{G}_{\rm C}) \end{pmatrix} \\
&= \begin{pmatrix}
 8 n \left(48 N^3+72 N^2+25 N+1\right) & 8 N \left(6 N^2+7 N+1\right) \\
 8 n \left(6 N^2+7 N+1\right) & 8 N (N+1) \\
\end{pmatrix}
\end{align}
where in the second line we have cast the result in terms of the mean photon number $N\equiv \langle {\rm SQV}|\hat{a}^\dag \hat{a} |{\rm SQV}\rangle$. Note that while the off-diagonal elements of the QFIM are non-zero, they are significantly smaller in the large $N$ limit suggesting the quantum and classical signatures of the single-mode gravitational self-interaction are highly distinguishable. Inverting this matrix gives:
\begin{align}
    \mathbf{F}_{\rm Q}^{-1} &=\frac{1}{96 N^2 (N+1)^2}  \begin{pmatrix}
 1 & -6 N-1 \\
 -6 N-1 & 24 N (2 N+1)+1 
\end{pmatrix} \,.
\end{align}
The top left element of this matrix gives Eq.~(7) in the main text.

\section{Proof that the the classical Fisher information from the Mach-Zehnder measurement scheme has similar scaling to the quantum Fisher information}
For a measurement scheme with an associated probability distribution $P_m$, the sensitivity is quantified by the (classical) Cram\'er-Rao bound (CRB) $\sqrt{M}\chi_{\rm Q}\geq [F^{\rm C}_{i,j}]^{-1}_{\rm Q,Q}$, where $F^{\rm C}_{i,j}=\sum_m \partial_{\chi_i}P_m\partial_{\chi_j}P_m/P_m$ is the \emph{classical} Fisher information matrix (CFIM). A measurement scheme is optimal if the CRB saturates the QCRB, i.e. if $[F^{\rm C}_{i,j}]^{-1}_{\rm Q,Q}=[F^{\rm Q}_{i,j}]^{-1}_{\rm Q,Q}$. As discussed in the main text, the signatures of quantum and classical gravity have orthogonal signatures in $P_m$, so the off-diagonal terms of the CFIM are zero. We now calculate the diagonal term to show that our measurement scheme has similar scaling to the QCRB. The Mach-Zehnder scheme includes a beamsplitter before and after the light enters the optical cavities,  and is therefore summarised by the unitary
\begin{align}
\hat{U} &= \exp\left[( -i \frac{\pi}{4}\left(\ahatd\bhat + \bhatd\ahat\right)\right]\exp\left[ i \chi_{\rm Q}\left(\ahatd\ahatd\ahat\ahat + \bhatd\bhatd\bhat\bhat \right)\right]\exp\left[( i\frac{\pi}{4}\left(\ahatd\bhat + \bhatd\ahat\right)\right] \notag\\
&= \exp\left[ i \chi_{\rm Q} \frac{1}{2}\left(\ahatd\ahatd\ahat\ahat + \bhatd\bhatd\bhat\bhat  +4\ahatd\ahat\bhatd\bhat - \ahatd\ahatd\bhat\bhat - \ahat\ahat \bhatd\bhatd \right)\right]
\end{align}
where the first and last exponentials represent the beam-splitting process, and the middle exponential represents the quantum gravity interaction. 
Therefore, the generator for our entire interaction (the gravitational self interaction and the beamsplitters) is 
\begin{align}
\hat{G} &=  \frac{1}{2}\left(\ahatd\ahatd\ahat\ahat + \bhatd\bhatd\bhat\bhat  +4\ahatd\ahat\bhatd\bhat - \ahatd\ahatd\bhat\bhat - \ahat\ahat \bhatd\bhatd \right) \, .
\end{align}
It is useful to break this down into $\hat{G} = \hat{G}_+ + \hat{G}_-$, with
\begin{align}
\hat{G}_+ &= \ahatd\ahatd\ahat\ahat + \bhatd\bhatd\bhat\bhat  +4\ahatd\ahat\bhatd\bhat ,
\end{align}
and
\begin{align}
\hat{G}_- &= - \ahatd\ahatd\bhat\bhat - \ahat\ahat \bhatd\bhatd.
\end{align}
We restrict ourselves to initial states of the form 
\begin{align}
|\Psi_0\rangle &= \sum_n c_n |n, n\rangle
\end{align}
of which two-mode squeezed vacuum is a subset. Following the method in Nolan \emph{et al}~\cite{Nolan:2017b}
we can calculate the classical Fisher information, $F_C$, by using
\begin{align}
d_H^2 &= \frac{\chi_{\rm Q}^2}{8} F_c
\end{align}
where $d_H^2$ is the Hellinger distance, given by
\begin{align}
d_H^2 &= 1 - \sum_{n_1, n_2} \sqrt{P_{n_1, n_2}(\chi_{\rm Q}) P_{n_1, n_2}(0)} \, ,
\end{align}
where $P_{n_1, n_2}(\chi_{\rm Q})$ is the probability of detecting $n_1$($n_2$) photons at detector 1(2), ie
\begin{align}
P_{n_1,n_2}(\chi_{\rm Q}) &= \abs{\langle n_1, n_2 | \hat{U}(\chi_{\rm Q}) |\Psi_0 \rangle}^2 . 
\end{align}
Now, given that $P_{n_1, n_2}(0) = \delta_{n_1, n}\delta_{n_2, n}|c_n|^2$, we can simplify this. Specifically, 
\begin{align}
d_H^2 &= 1 - \sum_n\sqrt{P_{n,n}(\chi_{\rm Q}) P_{n,n}(0)} \notag \\
&= 1 - \sum_n\sqrt{|\sum_m c_m \langle n,n |\hat{U}(\chi_{\rm Q})  |m,m\rangle |^2 P_{n,n}(0)} \, .
\end{align}
Noting that as $\hat{G}$ conserves the total number of photons,  $\langle n, n| \hat{G}|m, m\rangle = \langle n, n| \hat{G}|n,n\rangle \delta_{n,m}$, and therefore $\langle n, n| \hat{U}(\chi_{\rm Q})|m, m\rangle = \delta_{n,m}\langle n, n| \hat{U}(\chi_{\rm Q})|n, n\rangle$, $d_H^2$ simplifies to
\begin{align}
d_H^2 &= 1 - \sum_n\sqrt{|\langle n,n |\hat{U}(\chi_{\rm Q})  |n,n\rangle |^2 P^2_{n,n}(0)} \, .
\end{align}
Expanding $\hat{U}(\chi_{\rm Q})$ to second order in $\chi_{\rm Q}$ gives
\begin{align}
d_H^2 &= 1 - \sum_n P_{n,n}(0)\sqrt{|\langle n,n |\left(1 + i \chi_{\rm Q} \hat{G} - \frac{\chi_{\rm Q}}{2}\hat{G}^2 + \mathcal{O}\chi_{\rm Q}^3 \right)|n,n\rangle |^2 } \notag \\
&= 1 - \sum_n P_{n,n}(0)\sqrt{1 -  \chi_{\rm Q}^2 \left(\langle n,n |\hat{G}^2|n,n\rangle - \langle n,n |\hat{G}|n,n\rangle^2\right) + \mathcal{O}(\chi_{\rm Q}^3)} \notag \\
&\approx 1 - \sum_n P_{n,n}(0)\left(1 - \frac{\chi_{\rm Q}^2}{2}\left(\langle n,n |\hat{G}^2|n,n\rangle - \langle n,n |\hat{G}|n,n\rangle^2\right)\right) \notag \\
&=  \frac{\chi_{\rm Q}^2}{2}\sum_n P_{n,n}(0)\left(\langle n,n |\hat{G}^2|n,n\rangle - \langle n,n |\hat{G}|n,n\rangle^2\right) \, .
\end{align}
Therefore
\begin{align}
F_C &= \frac{8}{\chi_{\rm Q}^2} d_H^2 = 4\sum_n P_{n,n}(0)\left(\langle n,n |\hat{G}^2|n,n\rangle - \langle n,n |\hat{G}|n,n\rangle^2\right) \, .
\end{align}
We can simplify this further by noting that $\hat{G}_+|n,n\rangle = f(n)|n,n\rangle$, and $\hat{G}_- = \beta_1|n+2, n-2\rangle + \beta_2 |n-2, n+2\rangle$. As such, $\langle n, n| \hat{G}_-|n,n\rangle = 0$, $\langle n, n| \hat{G}_+^2|n,n\rangle = \langle n, n| \hat{G}_+|n,n\rangle^2$, and $\langle n, n| \hat{G}_+\hat{G}_-|n,n\rangle = \langle n, n| \hat{G}_-\hat{G}_+|n,n\rangle = 0$, and therefore
\begin{align}
\sum_n P_{n,n}(0)\left(\langle n,n |\hat{G}^2|n,n\rangle - \langle n,n |\hat{G}|n,n\rangle^2\right) &= \sum_n P_{n,n}(0)\langle n,n |\hat{G}_-^2|n,n\rangle \notag \\
&= \langle \hat{G}_-^2\rangle \, ,
\end{align} 
where we have used the fact that $\hat{G}_-$ doesn't couple between states with different total photon numbers, and therefore $\sum_{m,n} c_m^* c_n \langle m,m| \hat{G}_-^2|n,n\rangle = \sum_{n} |c_n|^2 \langle n,n| \hat{G}_-^2|n,n\rangle$ in the second line. For a two-mode squeezed vacuum state, this gives $F_C = 192 N^4$, to leading order in $N = \langle(\ahatd\ahat + \bhatd\bhat)\rangle$. We can compare this to the quantum Fisher information
\begin{align}
F_Q &= 4 \mathrm{Var}(G) = 320 N^4 \approx 1.6 F_C
\end{align}
to leading order in $N$. As such, the Mach-Zehnder measurement scheme is close to saturating the QCRB, and importantly, has the same scaling with the total number of photons. 

\bibliography{bib_Qgrav_simon3}

\end{document}